\documentclass[epjST]{svjour}

\newcommand{\ba}{\begin{align}}
\newcommand{\ea}{\end{align}}    
\newcommand{\be}{\begin{equation}}
\newcommand{\ee}{\end{equation}}
\newcommand{\gdualn}[1]{\overset{\:{}^{{}^{\boldsymbol{\neg}}}}{\smash[t]{#1}}} %Elko dual
\usepackage{graphicx}
\usepackage{graphics}
\usepackage{amssymb}
\usepackage{amsmath}
\usepackage{braket}
\usepackage{epstopdf}

\begin{document}
\title{Mass dimension one fermions at 1-loop}
%\title{Mass dimension one fermions quantum field theory at 1-loop}
\author{G. P. de Brito\inst{1} \fnmsep\thanks{\email{gpbrito@cbpf.br}} \and J. M. Hoff da Silva\inst{2}\fnmsep\thanks{\email{julio.hoff@unesp.br}} \and Vahid Nikoofard\inst{1,3}\fnmsep\thanks{\email{vahid@fat.uerj.br}} }
\institute{Centro Brasileiro de Pesquisas F\'isicas (CBPF),\\Rua Dr. Xavier Sigaud 150, Urca, Rio de Janeiro, Brazil, CEP 22290-180. \and Departamento de F\'isica e Qu\'imica, Universidade Estadual Paulista (UNESP),\\Av. Dr. Ariberto Pereira da Cunha, 333, Guaratinguet\'a, SP, Brazil. \and Departamento de Matem\'atica, F\'isica e Computa\c{c}\~ao,\\Faculdade de Tecnologia, Universidade do Estado do Rio de Janeiro (UERJ),\\Rodovia Pres. Dutra, Km 298, Polo Industrial, Resende, RJ, Brazil, CEP 27537-000.}
\abstract{
We investigate 1-loop contributions to the 2- and 4-point correlation functions within a theory for fermions with mass dimension one with a simple quartic self-interaction term. The 1-loop divergence appearing the self-energy function can be consistently absorbed by renormalization of the mass parameter. In the case of the (1PI) 4-point correlation function, there is a surprising cancellation of divergences coming from different diagrams, leading to a finite 1-loop result. The inclusion of a complete basis of four-fermion interactions is discussed. 
} %end of abstract
\maketitle

%\tableofcontents

\section{Introduction} 
\label{intro}

Mass dimension one spinor fields are constructed from the very realm of quantum field theory as candidates to describe dark matter \cite{Ahl,aac}. These fields are completely neutral under known gauge interactions and are built not to furnish an eigeinspinor relation under the action of parity operator. This last aspect allows the fields to have canonical mass dimension one, while the former may be faced as the responsible for its darkness. The neutrality condition is imposed on the expansion coefficients level by requiring these spinors to be eigeinspinors of the charge conjugation operator. The spinors resulting from this procedure are the so called Elko. Its is fairly trivial to see that, despite its constructed darkness, some couplings are still in order and some of them may lead to specific signals in high energy experiments \cite{mac}. 

It was demonstrated quite recently that these fields fulfill the necessary and sufficient conditions to frame their one particle states into an unusual class found by Wigner \cite{JR} when extending the irreducible representations of the Poincar\`e group to encompass discrete symmetries \cite{Wig}. 

In this paper we investigate the quantum effects of the mass and self interaction terms by inspecting its $1-$loop behavior. Loop corrections were previously considered in Ref. \cite{CYM}, were the authors have investigated a constraint in the mass parameter in order to ensure unitarity, as well as a Yukawa interaction term. The whole analysis was performed for the first version of the theory, where a broken Lorentz term were present in the spin sums, with effects in the propagator and other quantum correlators \cite{old}. Here we shall make use of perturbative techniques in order to study the consistency of the theory {\it per si}, i. e., analyzing purely self-interaction terms associated with mass dimension one fermions, without any other field. Besides, here we study the recent version of the theory in which a judicious investigation of the spinor dual theory leads to a Lorentz invariant formulation.  

This paper is organized as follows: in Section 2 we present a brief review on the formulation of a quantum field theory for a fermion with mass dimension one. In Section 3 we investigate 1-loop quantum corrections coming from a simple four-fermion interaction term. Section 4 is devoted to a brief discussion about the inclusion of additional quartic self-interaction terms. Finally, in Section 5 we present our concluding remarks.

\section{Some remarks on mass dimension one spinors}

In this section we review some basic aspect regarding the construction of a quantum field in terms of Elkos. 
In the context of usual canonical quantization, we can define a quantum field $\eta(x)$ by using the eigenvectors of the charge conjugation operator $\lambda^A_\sigma(\bf{p})$ and $\lambda^S_\sigma(\bf{p})$ as follows
\begin{align}
\eta(x)=\int \frac{d^3 \textbf{p} }{(2 \pi )^3}\frac{1}{\sqrt{4 m E(\textbf{p})}} \sum_{\sigma } \left(\!\!\frac{}{} a_{\sigma}(\textbf{p}) \lambda_{\sigma }^S (\textbf{p}) e^{-i p \cdot x} + b_{\sigma }^{\dagger}(\textbf{p})  \lambda_{\sigma }^A(\textbf{p}) e^{i p\cdot x}\right).
\end{align}
The dual field associated with $\eta(x)$ may be cast in the following form
\begin{align}
\gdualn{\eta} (x)=\int \frac{d^3 \textbf{p}} {(2 \pi )^3}\frac{1}{\sqrt{4  m E(\textbf{p})}} \sum_{\sigma} \left( a_{\sigma}^\dagger(\textbf{p}) \gdualn{\lambda}_{\sigma }^S(\textbf{p}) e^{i p\cdot x} + b_{\sigma }(\textbf{p}) \gdualn{\lambda}_{\sigma}^A(\textbf{p}) e^{-i p\cdot x}\right).
\end{align}
All the subtleties associated with a consistent construction of the Elko dual can be found, for instance, in Ref. \cite{Ahl}. Here we shall just pinpoint the quantum aspects necessary to the rest of the paper.  

As the Elkos obey the Fermi-Dirac statistics, the creation and annihilation operators $a_{\alpha}(\textbf{p})$ and $b_{\alpha}(\textbf{p})$ satisfy the anti-commuting algebra
\begin{align}\label{anti-commutation}
\left\{a_{\sigma}(\textbf{p}),a_{\sigma '}^\dagger (\textbf{q})\right\}=(2 \pi )^3 \delta ^3 (p-q) \delta _{\sigma \sigma '}=\left\{b_{\sigma}(\textbf{p}),b_{\sigma '}^\dagger (\textbf{q})\right\} ,
\end{align}
with the others anti-commutators being zero. 

Furthermore, the vacuum state $\ket{0}$ is defined in the usual way in terms of creation and annihilation operators
\begin{align}
a_{\sigma }(\textbf{p})\ket{0} = b_{\sigma}(\textbf{p})\ket{0} = 0
\qquad \textmd{and} \qquad 
\bra{0}a_{\sigma}^\dagger(\textbf{p}) = \bra{0}b_{\sigma}^\dagger(\textbf{p}) = 0.  
\end{align}
Taking into account the appropriate definition for the Elko dual, the expansion coefficients satisfy the following completeness relations 
\begin{align}
\sum_\sigma \lambda_{\sigma}^A(p^\mu) \gdualn{\lambda}_{\sigma}^A(p^\mu) = -2 m \mathbb{I}_4 	
\qquad \textmd{and} \qquad
\sum_\sigma \lambda_{\sigma}^S(p^\mu) \gdualn{\lambda}_{\sigma}^S(p^\mu) = 2 m \mathbb{I}_4	\,.
\end{align}

The time-ordered 2-point correlation function can be calculated using the expansion of the Elko in terms of creation and annihilation operators, rendering the following result
\begin{align}\label{propagator}
\bra{0}T \eta(x)\gdualn{\eta}(y)\ket{0}=\int_{C_F}\frac{d^4 p}{(2 \pi)^4}  \, D_{\textmd{Elko}}(p)\, 
e^{-i p\cdot (x-y)},
\end{align}
where $C_F$ stands for the usual Feynman's integration contour, and the momentum space propagator is given by
\begin{align}\label{Tree-Propagator}
D_{\textmd{Elko}}(p) = \frac{i}{p^2-m^2+i\epsilon} \mathbb{I}_4 \,.
\end{align}
It is straightforward to verify that the remaining 2-point correlators vanish 
\begin{align}
\bra{0}T \gdualn\eta(x)\gdualn{\eta}(y)\ket{0} =\bra{0}T \eta(x)\eta(y)\ket{0} = 0 \,.    
\end{align}
The main aspect that we should emphasis is the fact that the Elko free-propagator has the same structure, apart from a multiplicative identity in the Clifford algebra, as the usual free-propagator of a (complex) scalar field with dynamics specified by the Klein-Gordon Lagrangian. This fact leads to the unusual canonical mass dimension of the field, despite of its fermionic nature.

Higher order correlators can be computed in terms of the 2-point function via Wick theorem for interaction-free theories, namely
\begin{align}
&\bra{0} T\eta _{\alpha _1}(x_1)\gdualn{\eta}_{\beta _1}(y_1)\ldots \eta_{\alpha _n}(x_n)\gdualn{\eta}_{\beta _n}(y_n)\ket{0} = \\
&\qquad =\underset{i_1,\cdots,i_n=1}{\overset{n}{\sum }} \varepsilon ^{i_1 \text{$\ldots $i}_n} \bra{0}T \eta_{\alpha_1}(x_1)\gdualn{\eta}_{\beta_{i_1}}(y_{i_1})\ket{0}
\cdots 
\bra{0}T \eta_{\alpha_n}(x_n)\gdualn{\eta}_{\beta_{i_n}}(y_{i_n})\ket{0} \, ,  \nonumber
\end{align}
where the $\alpha$'s and $\beta$'s indicate the spinor indices. The Levi-Civita symbol accounts for the appropriate signs in accordance with the anti-commutation relations in \eqref{anti-commutation}.

For the sake of completeness, let us mention that in the presence of interactions, the correlation functions can be computed in terms of the free-field correlators by means of the textbook formula
\begin{align}
&\bra{\Omega} T\,\eta _{\alpha _1}(x_1)\gdualn{\eta}_{\beta _1}(y_1)\ldots \eta_{\alpha _n}(x_n)\gdualn{\eta}_{\beta _n}(y_n)\ket{\Omega} =\nonumber \\
&\qquad=\frac{\bra{0} T\,\eta _{\alpha _1}(x_1)\gdualn{\eta}_{\beta _1}(y_1)\ldots \eta_{\alpha _n}(x_n)\gdualn{\eta}_{\beta _n}(y_n) e^{i \int d^4x \mathcal{L}_{\textmd{int}} [\eta,\gdualn{\eta}]} \ket{0} }{\bra{0} T\, e^{i \int d^4x \mathcal{L}_{\textmd{int}} [\eta,\gdualn{\eta}]} \ket{0}} \,,
\end{align}
with $\mathcal{L}_{\textmd{int}} $ denoting the interacting part of the Lagrangian describing the Elko dynamics and $\ket{\Omega}$ being the vacuum state for the interacting theory.

\section{Four-fermion interaction of the mass dimension one spinor}

The canonical mass dimension of the Elko leads to the possibility of four-fermion interaction terms which are not compatible with the Dirac (or Weyl/Majorana) spinors within a perturbatively renormalizable QFT in four space-time dimensions. In this essay, we present some results for 1-loop computations in a QFT for an Elko with a single quartic interaction term given by
\begin{align}\label{4-interaction}
\mathcal{L}_{\textmd{int}} = -\frac{g}{2}(\gdualn{\eta}\eta\,)^2.
\end{align}
Since the self-coupling $g$ has canonical mass dimension zero, this interaction is classified as power-counting renormalizable. In fact, taking into account the tree-level propagator given by \eqref{Tree-Propagator} along with the quartic interaction in \eqref{4-interaction}, it is not difficult to see that a Feynman diagram $\mathcal{G}$ (with $n_E$ external legs) exhibits superficial degree of divergence given by
\begin{align}
\omega_{\textmd{Elko}}(\mathcal{G}) = 4 - n_E \,.
\end{align}
Note that $\omega_{\textmd{Elko}}(\mathcal{G})$ is similar to the superficial degree of divergence associated with a scalar theory with quartic self-interaction, namely $\omega_{\textmd{scalar}}(\mathcal{G}) = 4 - n_E$.
In contrast, the superficial degree of divergence for a diagram $\mathcal{G}$ within a 4-dimensional QFT for a Dirac fermion with quartic interaction like $\mathcal{L}_{\textmd{int}}^{\textmd{Dirac}} \sim (\bar{\psi} \psi)^2$ is given by $\omega_{\textmd{Dirac}}(\mathcal{G}) = 4 + 2 n_V - 3n_E/2$, which increase with the number of vertices $n_V$.

In this paper we focus in the structure of divergences arising from 1-loop diagrams computed within a QFT for the Elko with quartic self-interaction given by \eqref{4-interaction}. Using $\omega_{\textmd{Elko}}(\mathcal{G})$ one can easily see that the superficially divergent diagrams are those with 2- and 4-external legs. The 1-loop relevant diagrams have been shown in Figs. \ref{Fig_2pt} and \ref{Fig_4pt}. 

\subsection{Self-energy contribution}

\begin{figure}[htb!]
	\begin{center}
		\includegraphics[width=4.5cm]{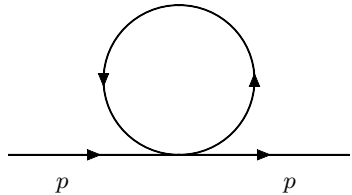}
		\put(-110,-10){$p$}
		\put(-25,-10){$p$}
		\caption{1-loop contribution to the self-energy $\Sigma_{\textmd{Elko}}(p^2)$ within a QFT describing an Elko with quartic interaction \eqref{4-interaction}.}
		\label{Fig_2pt}
	\end{center}
\end{figure}

We start with the 1-loop contribution to the self-energy $\Sigma_{\textmd{Elko}}(p^2)$. As usual, we have defined $\Sigma_{\textmd{Elko}}(p^2)$ in terms of the connected 2-point function, namely
\begin{align}\label{propagator_full}
\bra{\Omega}T \eta(x)\gdualn{\eta}(y)\ket{\Omega}_{\textmd{conn.}}
=\int_{C_F}\frac{d^4 p}{(2 \pi)^4}  \, \frac{i}{p^2 - m^2 -\Sigma_{\textmd{Elko}}(p^2) + i \,\epsilon} 
e^{-i p\cdot (x-y)} \, \mathbb{I}_4 \, .
\end{align}
The 1-loop contribution is given by tadpole diagram represented in Fig. \ref{Fig_2pt} and can be translated into the following analytical result
\begin{align}\label{Elko_Self_Energy}
\Sigma_{\textmd{Elko}}^{\textmd{1-loop}} = -3 \,g \int \frac{d^4k}{(2\pi)^4} \frac{i}{k^2-m^2+i\epsilon} \, .
\end{align}

It is instructive to compare this result with the corresponding one coming from a scalar field theory with self-interaction $\lambda \phi^4$. In this case, the usual textbook formula for the self-interaction is given by
\begin{align}
\Sigma_{\textmd{Scalar}}^{\textmd{1-loop}} = \frac{\lambda}{2} \int \frac{d^4k}{(2\pi)^4} \frac{i}{k^2-m_\phi^2+i\epsilon} \, ,
\end{align}
where $m_\phi$ correspond to the mass of the scalar field $\phi$.

The loop integral appearing in $\Sigma_{\textmd{Elko}}^{\textmd{1-loop}}$ is exactly the same as the one in $\Sigma_{\textmd{Scalar}}^{\textmd{1-loop}}$. This is a direct consequence of the similarity between the Elko and Klein-Gordon tree-level propagators. On the other hand, a crucial difference between $\Sigma_{\textmd{Elko}}^{\textmd{1-loop}}$ and $\Sigma_{\textmd{Scalar}}^{\textmd{1-loop}}$, reflecting the fermionic nature of the Elko field, is the negative sign in \eqref{Elko_Self_Energy} as a consequence of the fermionic loop in Fig. \ref{Fig_2pt}.

The 1-loop contribution to the self-energy can be computed using dimensional regularization and results the expression
\begin{align}
\Sigma_{\textmd{Elko}}^{\textmd{1-loop}} = 
-\frac{3 g \,\mu^{-\varepsilon} \, m^2}{16\pi^2} \bigg[ \!-\frac{2}{\varepsilon} + \gamma_\textmd{E} - 1 + \ln \bigg( \frac{m^2}{4\pi \mu^2} \bigg) \bigg] ,
\end{align}
where $\gamma_\textmd{E}$ corresponds to the Euler-Mascheroni constant and $\mu$ is the typical mass parameter in the dimensional regularization procedure.

\subsection{4-point correlation function}

Now we turn our attention to the 1-loop contribution to the 4-point correlation function. We focus on the leading radiative correction to the vertex function $i\Gamma^{(4)}_{\textmd{Elko}}$. In order to make it more precise, the vertex function is defined by summing up the 1PI contributions to the 4-point correlator $\bra{\Omega} T \,\eta_{\alpha_1}(x_1) 
\gdualn{\eta}_{\beta_1}(y_1) \eta_{\alpha_2}(x_2) \gdualn{\eta}_{\beta_2}(y_2) \ket{\Omega}$ with amputated external propagators. 
\begin{figure}[htb!]
	\begin{center}
		\includegraphics[width=12cm]{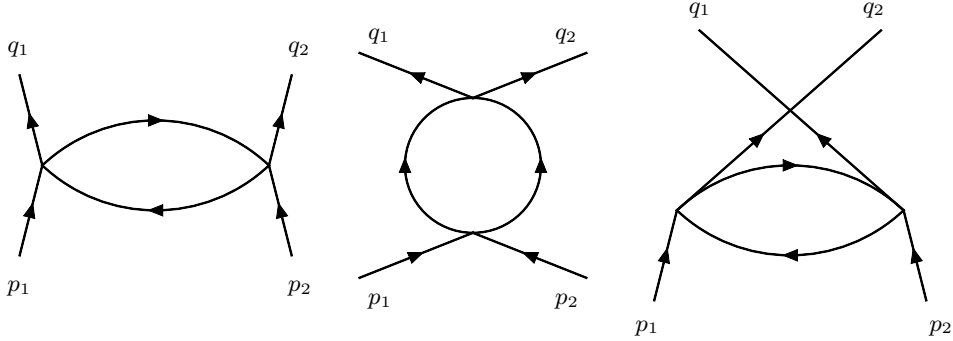}
		\put(-345,5){$p_1$}
		\put(-240,5){$p_2$}
		\put(-345,95){$q_1$}
		\put(-240,95){$q_2$}
		%%%%%%%%%%%%%%%%%%%%%%%%%%
		\put(-210,0){$p_1$}
		\put(-140,0){$p_2$}
		\put(-210,100){$q_1$}
		\put(-140,100){$q_2$}
		%%%%%%%%%%%%%%%%%%%%%%%%
		\put(-110,-10){$p_1$}
		\put(-0,-10){$p_2$}
		\put(-90,110){$q_1$}
		\put(-25,110){$q_2$}
		%%%%%%%%%%%%%%%%%%%%%%
		\caption{1PI diagrams contributing to the 1-loop correction to the vertex function $i\Gamma^{(4)}_\textmd{Elko}$. Following usual power-counting arguments, each diagram diverges logarithmically.}
		\label{Fig_4pt}
	\end{center}
\end{figure}

The 1-loop contribution to $i\Gamma^{(4)}_\textmd{Elko}$ is obtained by adding the 1PI diagrams represented in Fig. \ref{Fig_4pt}, which can be expressed (in momentum space) in the following way
\begin{align}\label{1-loop_vertex_Elko}
\left[i\Gamma^{(4)}_\textmd{Elko}(p_1,q_1,p_2,q_2)\right]^{\textmd{1-loop}}_{\alpha_1\beta_1\alpha_2\beta_2} &= \,
g^2 \, (2 \,\delta_{\alpha_1 \beta_1} \delta_{\alpha_2 \beta_2} + \delta_{\alpha_1 \beta_2} \delta_{\alpha_1 \beta_2}) \,\mathcal{I}(p_1-q_1) \nonumber \\
&\frac{}{}-g^2 \,( \delta_{\alpha_1 \beta_1} \delta_{\alpha_2 \beta_2} - \delta_{\alpha_1 \beta_2} \delta_{\alpha_1 \beta_2}) \,\mathcal{I}(p_1+p_2)  \nonumber \\
&\frac{}{}-g^2 \, (\delta_{\alpha_1 \beta_1} \delta_{\alpha_2 \beta_2} + 2\,\delta_{\alpha_1 \beta_2} \delta_{\alpha_1 \beta_2}) \,\mathcal{I}(p_1-q_2) \,,
\end{align}
with implicit momentum conservation $p_1+p_2-q_1-q_2 = 0$. The relevant 1-loop integral $\mathcal{I}(p)$ takes the form
\begin{align}
\mathcal{I}(p) = \int \frac{d^4 k}{(2\pi)^4} \frac{i}{k^2 - m^2 + i \epsilon} \frac{i}{(p-k)^2 - m^2 + i \epsilon}.
\end{align}
The similarities with the scalar field case persist in the sense that $\mathcal{I}(p)$ also appears as the relevant integral in the 1-loop contribution to the vertex function in the $\lambda \phi^4$-theory. 
For the sake of comparison, we recall that 
\begin{align}\label{1-loop_vertex_Scalar}
i\Gamma^{(4)}_\textmd{Scalar}(p_1,p_2,p_3,p_4)\Big|_{\textmd{1-loop}} = \,
-\frac{1}{2}\lambda^2 \Big( \mathcal{I}(p_1+p_2)+ \mathcal{I}(p_1+p_3) +\mathcal{I}(p_1+p_4) \Big) \,.
\end{align}
However, once we compare \eqref{1-loop_vertex_Elko} and \eqref{1-loop_vertex_Scalar} we note that, different from the 2-point case, the 1-loop contribution to $i\Gamma^{(4)}_\textmd{Elko}$ cannot be obtained by multiplying the $i\Gamma^{(4)}_\textmd{Scalar}|_{\textmd{1-loop}}$ with some overall ``tensorial structure''. This is a consequence of different combinations of $\delta$'s arising from different diagrams in Fig. \ref{Fig_4pt}.

The structure of 1-loop divergences in each diagram contributing to $[i\Gamma^{(4)}_\textmd{Elko}]^{\textmd{1-loop}}$ takes exactly the same form as in the $\lambda \phi^4$-theory. As a matter of fact, within dimensional regularization the 1-loop integral $\mathcal{I}(p)$ is given by
\begin{align}\label{dim_reg_Int}
\mathcal{I}(p) = \frac{i \mu^{-\varepsilon}}{16 \pi^2} \bigg[\! -\frac{2}{\varepsilon} 
+  \gamma_{\textmd{E}} + \int_0^1 dx \, \ln\left( \frac{m^2 - x(1-x)p^2}{4\pi \mu^2} \right)  \bigg] .
\end{align}

A surprising result, however, arises when we replace the regularized integral $\mathcal{I}(p)$ into the 1-loop contribution to the vertex function $i\Gamma^{(4)}_\textmd{Elko}$. In this case, the momentum independent part in \eqref{dim_reg_Int},
\begin{align}
\frac{i \mu^{-\varepsilon}}{16 \pi^2} \left( \! -\frac{2}{\varepsilon} +  \gamma_{\textmd{E}} \right) \,,
\end{align}
cancels out when we put together contributions from the different diagrams in Fig. \ref{Fig_4pt}. As a consequence, the resulting 1-loop contribution to vertex $[i\Gamma^{(4)}_\textmd{Elko}]$ turns out to be finite. Using the regularized expression \eqref{dim_reg_Int}, we find
\begin{align}
&\left[i\Gamma^{(4)}_\textmd{Elko}(p_1,q_1,p_2,q_2)\right]^{\textmd{1-loop}}_{\alpha_1\beta_1\alpha_2\beta_2} =  \nonumber\\
&\frac{}{}= \frac{i g^2 \mu^{-\varepsilon}}{16 \pi^2} \bigg[\, (2 \,\delta_{\alpha_1 \beta_1} \delta_{\alpha_2 \beta_2} + \delta_{\alpha_1 \beta_2} \delta_{\alpha_1 \beta_2})  \int_0^1 dx \, \ln\left( \frac{m^2 - x(1-x)(p_1-q_1)^2}{4\pi \mu^2} \right) \nonumber \\
&\frac{}{}- \,( \delta_{\alpha_1 \beta_1} \delta_{\alpha_2 \beta_2} - \delta_{\alpha_1 \beta_2} \delta_{\alpha_1 \beta_2}) \int_0^1 dx \, \ln\left( \frac{m^2 - x(1-x)(p_1+p_2)^2}{4\pi \mu^2} \right)  \nonumber \\
&\frac{}{}- \, (\delta_{\alpha_1 \beta_1} \delta_{\alpha_2 \beta_2} + 2\,\delta_{\alpha_1 \beta_2} \delta_{\alpha_1 \beta_2}) \int_0^1 dx \, \ln\left( \frac{m^2 - x(1-x)(p_1-q_2)^2}{4\pi \mu^2} \right) \bigg] 
\,. 
\end{align}
In particular, it is not difficult to check that this expression remains finite when we remove the regulator, i.e., in the limit $\varepsilon \to 0$. We emphasize that the cancellation of the 1-loop divergences observed here depend on a specific combination of the ``tensorial structures'' multiplying each diagram in Fig. \ref{Fig_4pt} and alternating signs arising from the anti-commuting nature of the Elko.

By now, it is not clear to us whether this cancellation is an accident from the 1-loop approximation or if there is a fundamental reason behind it. A direct test can be done by taking into account the 2-loops contributions to the vertex function, however, it goes beyond the scope of the present essay. We might also speculate the existence of some hidden symmetry which could provide a mechanism responsible for cancellation of UV divergences in more fundamental level. We shall return to this question in a future publication with further systematic analysis on the radiative corrections within a QFT for mass dimension one spinors.

\subsection{Renormalization}

Using the results of the previous section, we can cast the momentum space propagator, $G_{\textmd{Elko}}(p)$, with regularized 1-loop corrections in the following way
\begin{align}\label{propagator_1loop}
G_{\textmd{Elko}}^{\textmd{1-loop}}(p)
&= \frac{i}{p^2 - m^2 -\Sigma_{\textmd{Elko}}^{\textmd{1-loop}}(p^2) + i \,\epsilon}  \nonumber \\
&= \frac{i}{p^2 - m^2 + \frac{3 g \,\mu^{-\varepsilon} \, m^2}{16\pi^2} \left[ \!-\frac{2}{\varepsilon} + \gamma_\textmd{E} - 1 + \ln \left( \frac{m^2}{4\pi \mu^2} \right) \right]+ i \,\epsilon}  \,.
\end{align}
Further, taking into account \eqref{1-loop_vertex_Elko}, it is convenient to recast the vertex function (including the tree-level contribution) as follows
\begin{align}
&\left[i\Gamma^{(4)}_\textmd{Elko}(p_1,q_1,p_2,q_2)\right]_{\alpha_1\beta_1\alpha_2\beta_2} = 
- ig\, (\delta_{\alpha_1 \beta_1} \delta_{\alpha_2 \beta_2} 
- \delta_{\alpha_1 \beta_2} \delta_{\alpha_2\beta_1})  \nonumber\\\,
&\frac{}{}+ \frac{i g^2 \mu^{-\varepsilon}}{16 \pi^2} 
\Bigg\{\, \,\delta_{\alpha_1 \beta_1} \delta_{\alpha_2 \beta_2} 
\int_0^1 dx \, \left[ \ln\left( \frac{F(p_1-q_1;x)}{F(p_1+p_2;x)} \right)  -
\ln\left( \frac{F(p_1+p_2;x)}{F(p_1-q_1;x)} \right) \right]  \nonumber \\
&\frac{}{}+ \delta_{\alpha_1 \beta_2} \delta_{\alpha_2 \beta_1} 
\int_0^1 dx \, \left[ \ln\left( \frac{F(p_1+p_2;x)}{F(p_1-q_2;x)} \right)  -
\ln\left( \frac{F(p_1-q_2;x)}{F(p_1-q_1;x)} \right) \right] \Bigg\} 
+ \mathcal{O}(g^3)\,,
\end{align}
where we have defined $F(p;x)= m^2 - x(1-x)\,p^2$. Up to now, our results were expressed in terms of the bare parameters $m^2$ and $g$. We define the renormalized mass parameter as
\begin{align}
m_R^2 = m^2 -\frac{3 g \,\mu^\varepsilon\, m^2}{16\pi^2} \bigg[\! -\frac{2}{\varepsilon} + \gamma_\textmd{E} - 1  \bigg]  \,.
\end{align}
Since the 1-loop contribution to the vertex function is finite, $g_R$ is defined through a finite renormalization relation
\begin{align}
g_R = g \, \mu^\varepsilon ,
\end{align}
which basically accounts for the appropriate mass dimension of $g_R$ and the corresponding bare coupling $g$.
We emphasize that, as it happens in the $\lambda \phi^4$, 1-loop corrections to the propagator associated with the Elko do not require wave function renormalization.
By expressing $G_{\textmd{Elko}}^{\textmd{1-loop}}(p)$ and $i\Gamma^{(4)}_\textmd{Elko}(p_1,q_1,p_2,q_2)$ in terms of renormalized coupling and taking the limit $\varepsilon \to 0$, we arrive at the finite results 
\begin{subequations}
	\begin{align}
	G_{\textmd{Elko}}^{\textmd{1-loop}}(p)
	= \frac{i}{p^2 - m_R^2 + \frac{3 g_R  \, m_R^2}{16\pi^2}  \ln \left( \frac{m_R^2}{4\pi \mu^2} \right) + i \,\epsilon}  \,,
	\end{align}
	and
	\begin{align}
	&\left[i\Gamma^{(4)}_\textmd{Elko}(p_1,q_1,p_2,q_2)\right]_{\alpha_1\beta_1\alpha_2\beta_2} = 
	- ig_R\, (\delta_{\alpha_1 \beta_1} \delta_{\alpha_2 \beta_2} 
	- \delta_{\alpha_1 \beta_2} \delta_{\alpha_2\beta_1})  \\\,
	&\frac{}{}+ \frac{i g_R^2 }{16 \pi^2} 
	\Bigg\{\, \,\delta_{\alpha_1 \beta_1} \delta_{\alpha_2 \beta_2} 
	\int_0^1 dx \, \left[ \ln\left( \frac{F(p_1-q_1;x)}{F(p_1+p_2;x)} \right)  -
	\ln\left( \frac{F(p_1+p_2;x)}{F(p_1-q_1;x)} \right) \right]  \nonumber \\
	&\frac{}{}+ \delta_{\alpha_1 \beta_2} \delta_{\alpha_2 \beta_1} 
	\int_0^1 dx \, \left[ \ln\left( \frac{F(p_1+p_2;x)}{F(p_1-q_2;x)} \right)  -
	\ln\left( \frac{F(p_1-q_2;x)}{F(p_1-q_1;x)} \right) \right] \Bigg\} 
	+ \mathcal{O}(g_R^3)\,. \nonumber
	\end{align}
\end{subequations}

As usual, the running of the renormalized couplings $m_R^2$ and $g_R$, at 1-loop, can be obtained by demanding that $G_{\textmd{Elko}}^{\textmd{1-loop}}(p)$ and $i\Gamma^{(4)}_\textmd{Elko}(p_1,q_1,p_2,q_2)$, expressed in terms of renormalized quantities, should be independent of the arbitrary mass parameter $\mu$ introduced in the dimensional renormalization procedure. For the renormalized mass parameter we find
\begin{align}
\mu \frac{\partial m_R^2}{\partial \mu} \bigg|_{\textmd{1-loop}}= -\frac{3 \,m_R^2\,g_R}{8\pi^2}.
\end{align} 
Note that the opposite sign in comparison with the running of renormalized mass parameter associated with a scalar field within $\lambda \phi^4$-theory, namely $\mu \,\partial m_{\phi,R}^2/\partial \mu |_{\textmd{1-loop}}= m_{\phi,R}^2\,g_R/16\pi^2$. As a consequence of the 1-loop finiteness of the vertex function, the beta function associated with the renormalized coupling $g_R$ turns out to be zero
\begin{align}
\beta_{g_R}^{\textmd{1-loop}} \equiv 
\mu \frac{\partial g_R}{\partial \mu} \bigg|_{\textmd{1-loop}} = 0  \,.
\end{align}

\section{Steps beyond the single four-fermion interaction \label{general_int}}

Up to this point we have considered quantum corrections originating from the single quartic self-interaction $(\gdualn{\eta}\eta\,)^2$. Nevertheless, even if we restrict ourselves to self-interactions of the Elko, Lorentz covariance allows the inclusion additional quartic interaction terms obtained by scalar combinations of bi-linear structures such as $\gdualn{\eta} \gamma^5 \eta$, $\gdualn{\eta} \gamma^\mu \eta$, $\gdualn{\eta} \gamma^5\gamma^\mu \eta$, $\gdualn{\eta} \gamma^\mu \gamma^\nu \eta$, etc. In the usual case of four-fermion interactions with Dirac fermion, the complete basis of quartic self-interactions is given by the Fierz basis. Up to now, however, a complete Fierz basis for the Elko has not been constructed. In fact, due to non-trivial subtleties in the definition of the Elko dual \cite{Ahl}, the construction of such basis may not correspond to the mere replacement of Dirac spinors by Elko in the usual Fierz basis.

As a first step towards the inclusion of terms beyond the single self-interaction $(\gdualn{\eta}\eta\,)^2$, we present some results for 1-loop computations involving the generalized quartic interaction
\begin{align}\label{4-interaction_generalized}
\mathcal{L}_{\textmd{int}} = -\frac{1}{4} g_{\alpha_1 \beta_1,\alpha_2 \beta_2} \,
\gdualn{\eta}_{\alpha_1} \eta_{\beta_1} \gdualn{\eta}_{\alpha_2} \eta_{\beta_2},
\end{align}
with generalized self-coupling $g_{\alpha_1 \beta_1, \alpha_2 \beta_2}$ with spinorial indices $\alpha$'s and $\beta$'s. Note that
\begin{align} \label{anti-sym}
g_{\alpha_1 \beta_1, \alpha_2 \beta_2} = -g_{\alpha_2 \beta_1 ,\alpha_1 \beta_2} = -g_{\alpha_1 \beta_2 ,\alpha_2 \beta_1} \,,
\end{align}
as a consequence of anti-commuting nature of fermions. We emphasize that without further considerations on the generalized coupling $g_{\alpha_1 \beta_1, \alpha_2 \beta_2}$ the interaction Lagrangian in \eqref{4-interaction_generalized} resembles a scalar expression. In order to turn $\mathcal{L}_{\textmd{int}}$ into a Lorentz invariant expression, the generalized coupling should be expanded in terms of combinations involving $\delta_{\alpha\beta}$'s and $\gamma$-matrices, namely
\begin{align} \label{gen_coupling}
g_{\alpha_1 \beta_1, \alpha_2 \beta_2} &= 
g_1 \left[\delta_{\alpha_1\beta_1} \delta_{\alpha_2\beta_2} - \delta_{\alpha_1\beta_2} \delta_{\alpha_2\beta_1}\right] \nonumber  \\
&\,+g_2 \left[(\gamma^\mu)_{\alpha_1\beta_1} (\gamma_\mu)_{\alpha_2\beta_2} - (\gamma^\mu)_{\alpha_1\beta_2} (\gamma_\mu)_{\alpha_2\beta_1} \right] + \, \cdots \, .
\end{align}
In principle, one can adjust the coupling $g$'s in order to cast $\mathcal{L}_{\textmd{int}}$ in terms of an appropriate Fierz basis. We note that the first term in the \eqref{gen_coupling} corresponds to the single quartic interaction \eqref{4-interaction} discussed in the previous section. In order to extract broad conclusions regarding quartic self-interaction for the Elko, the only assumption regarding $g_{\alpha_1 \beta_1 \alpha_2 \beta_2} $ that we are going to take into account is the anti-symmetric property given by \eqref{anti-sym}.

Recomputing the 1-loop contribution to the self-energy and vertex function, now including the generalized quartic interaction \eqref{4-interaction_generalized}, we find the following results
\begin{subequations}
	\begin{align}\label{Elko_Self_Energy_generalized}
	[\Sigma_{\textmd{Elko}}^{\textmd{1-loop}}]_{\alpha\beta} = - \,g_{\alpha\beta,\theta\theta} \int \frac{d^4k}{(2\pi)^4} \frac{i}{k^2-m^2+i\epsilon} \, ,
	\end{align}
	\begin{align}\label{1-loop_vertex_Elko_generalized}
	&\left[i\Gamma^{(4)}_\textmd{Elko}(p_1,q_1,p_2,q_2)\right]^{\textmd{1-loop}}_{\alpha_1\beta_1\alpha_2\beta_2} =  \,g_{\alpha_1\beta_1,\rho\theta} \, g_{\theta\rho,\alpha_2\beta_2} \,\mathcal{I}(p_1-q_1) \nonumber \\
	&\qquad\,- \frac{1}{2}g_{\alpha_1\rho,\alpha_2\theta} \, g_{\rho\beta_1,\theta\beta_2}
	\,\mathcal{I}(p_1+p_2) 
	- \,g_{\alpha_1\beta_2,\rho\theta} \, g_{\theta\rho,\alpha_2\beta_1} \,\mathcal{I}(p_1-q_2) \,.
	\end{align}
\end{subequations}
The relevant 1-loop integrals appearing both in the self-energy and vertex functions are the same as in the simple case of self-interaction given by \eqref{4-interaction}. The difference with respect to the previous case relies on the appearance of different ``tensorial structures'' associated with the generalized self-coupling $g_{\alpha_1 \beta_1, \alpha_2 \beta_2}$.

The integral present in the 1-loop contribution self-energy diagram produces a quadratic divergence which one has to absorb by renormalization of the mass parameter. However, in order to make this renormalization procedure viable, the divergence has to be proportional to the identity matrix in the Clifford algebra. Since the 1-loop contribution has a global factor $g_{\alpha \beta, \theta\theta}$, 1-loop renormalization requires the following constraint on the generalized coupling
\begin{align}
g_{\alpha \beta, \theta\theta} \propto \delta_{\alpha\beta} \,.
\end{align}

In the case of the 1-loop correction to the vertex function, each contribution has a logarithmic divergence arising from $\mathcal{I}(p)$. Using dimensional regularization we find
\begin{align}
&\left[i\Gamma^{(4)}_\textmd{Elko}(p_1,q_1,p_2,q_2)\right]^{\textmd{1-loop}}_{\alpha_1\beta_1\alpha_2\beta_2} =
\frac{i \mu^{-\varepsilon}}{8\pi^2 \,\varepsilon} \,\bigg( g_{\alpha_1\beta_1,\rho\theta} \, g_{\theta\rho,\alpha_2\beta_2}  \nonumber \\
&\qquad - \frac{1}{2}g_{\alpha_1\rho,\alpha_2\theta} \, g_{\rho\beta_1,\theta\beta_2}
- \,g_{\alpha_1\beta_2,\rho\theta} \, g_{\theta\rho,\alpha_2\beta_1}  \bigg)  + \,\textmd{Finite Contributions} \, .
\end{align} 
As one can see, the finite 1-loop result observed in the previous section does not hold anymore. In principle one can achieve 1-loop fineness by imposing the additional constraint
\begin{align}
g_{\alpha_1\beta_1,\rho\theta} \, g_{\theta\rho,\alpha_2\beta_2}  - \frac{1}{2}g_{\alpha_1\rho,\alpha_2\theta} \, g_{\rho\beta_1,\theta\beta_2}
 - \,g_{\alpha_1\beta_2,\rho\theta} \, g_{\theta\rho,\alpha_2\beta_1}  = 0 \,.
\end{align} 
By now it is not completely clear to us whether or not it is possible to find solutions to this equation besides the one corresponding to the simple interaction in \eqref{4-interaction}.

\section{Concluding remarks}

In this paper we have presented a first endeavor towards the understanding of perturbative quantum corrections generated by self-interaction of fermions with mass dimension one. In particular, taking into account the simplest quartic interaction term constructed with the Elko, $(\gdualn{\eta}\eta\,)^2$, we have shown some explicit results for 1-loop corrections to the self-energy and 4-vertex function. In the first case, the divergent contribution, expected by power counting arguments, requires the renormalization of the mass parameter. Regarding the 1-loop contribution to the 4-vertex function we have found a surprising cancellation of 1-loop the logarithmic divergences once we put together all the 1PI diagrams contributing to the 4-point correlation function up to the approximation under investigation. Since the theory for mass dimension one fermions has the same power counting of the usual scalar $\lambda \phi^4$-theory, we compare their structure of 1-loop corrections.
 
The whole work is an attempt to further explore perturbative quantization for a theory of mass dimension one fermions beyond the regular general power counting arguments. The obtained results open interesting questions to be addressed in the future. The first (and natural) question is about the persistence finite results to 4-point function once we include higher loop contributions. A direct test is the explicit computation of the 2-loops diagrams contributing to the 4-point function. If the finite result survives the inclusion of next-order contributions, an interesting approach would be the search for a hidden symmetry that could provide a mechanism for the cancellation of divergences in the 4-point function.  

Another point that deserves further investigation is the inclusion of additional four-fermion interactions constructed from possible (scalar) combinations of bi-linear structures associated with mass dimension one fermions. As we have discussed in section \eqref{general_int}, the quest for a complete set of four-fermion interactions motivates the search for Fierz basis associate with mass dimension one fermions. Furthermore, we have shown that the cancellation of 1-loop divergences obtained with a single quartic interaction cannot be trivially extended to generalized four-fermion interactions. 

\section*{Acknowledgments}
We acknowledge enlightening discussions with J. Helay\"el-Neto and S.A. Dias. GPB is grateful for the support by CNPq (Grant no.~142049/2016-6) and thanks the DFQ Unesp-Guaratinguet\'a for hospitality. JMHS thanks to CNPq (Grant no. 303561/2018-1) for partial financial support. VN would like to thank COSMO-CBPF for the hospitality.

%\begin{figure}
%\resizebox{0.75\columnwidth}{!}{%
%  \includegraphics{fig1.eps} }
%\caption{Please write your figure caption here.}
%\end{figure}
%\begin{table}
%\caption{Please write your table caption here.}
%\begin{tabular}{lll}
%\hline\noalign{\smallskip}
%\end{tabular}
%\end{table}
%


\begin{thebibliography}{}
\bibitem{Ahl} D. Ahluwalia, \emph{Mass Dimension One Fermions}, Cambridge University Press, London (2019). 

\bibitem{aac} D. Ahluwalia, \emph{The theory of local mass dimension one fermions of spin one half}, Adv.Appl.Clifford Algebras {\bf 27} (2017) 2247; D. Ahluwalia, \emph{Evading Weinberg's no-go theorem to construct mass dimension one fermions: Constructing darkness}, Europhys. Lett. {\bf 118} (2017) 60001.

\bibitem{mac} A. Alves, M. Dias, F. de Campos, L. Duarte, J. M. Hoff da Silva, \emph{Constraining Elko Dark Matter at the LHC with Monophoton Events},  Europhys. Lett. {\bf 121} (2018) 31001; A. Alves, M. Dias, F. de Campos, J. M. Hoff da Silva, \emph{Searching for Elko dark matter spinors at the CERN LHC}, Int. J. Mod. Phys. A {\bf 30} (2015) 1550006; M. Dias, F. de Campos, J. M. Hoff da Silva, \emph{Exploring Elko typical signature}, Phys. Lett. B {\bf 706} (2012) 352.

\bibitem{JR} J. M. Hoff da Silva and R. J. Bueno Rogerio, \emph{Massive spin one-half one particle states for the mass dimension one fermions}, to appear in Europhys. Lett [arXiv:1908.00458 [hep-th]].

\bibitem{Wig} E. P. Wigner, \emph{Unitary representations of the inhomogeneous Lorentz group including reflections}, in Group theoretical concepts and methods in elementary particle physics, Lectures of the Istanbul summer school of theoretical physics (1962), F. Guersey, ed., pp. 37-80. Gordon and Breach, New York, (1964).

\bibitem{CYM} C.-Y. Lee and M. Dias, \emph{Constraints on mass dimension one fermionic
	dark matter from the Yukawa interaction}, Phys. Rev. D {\bf 94} (2016) 065020. 

\bibitem{old} D. Ahluwalia and D. Grumiller, \emph{Spin half fermions with mass dimension one: Theory, phenomenology, and dark matter}, JCAP {\bf 0507} (2005) 012; D. Ahluwalia and D. Grumiller, \emph{Dark matter: A spin one half fermion field with mass dimension one?}, Phys. Rev. D {\bf 72} (2005) 067701; D. Ahluwalia, C.-Y. Lee, D. Schritt, \emph{Elko as self-interacting fermionic
	dark matter with axis of locality}, Phys. Lett. B {\bf 687} (2010) 248; D. Ahluwalia, C.-Y. Lee, D. Schritt, \emph{Self-interacting Elko dark matter with an axis of locality}, Phys. Rev. D {\bf 83} (2011) 065017.


\end{thebibliography}
\end{document}